\documentclass[10pt,prl,twocolumn,showpacs]{revtex4-1}
\usepackage{graphicx,amsmath,amssymb}
\usepackage{hyperref}
\begin{document}

\def\nn{\nonumber}
\def\kc#1{\left(#1\right)}
\def\kd#1{\left[#1\right]}
\def\ke#1{\left\{#1\right\}}
\renewcommand{\Re}{\mathop{\mathrm{Re}}}
\renewcommand{\Im}{\mathop{\mathrm{Im}}}
\renewcommand{\b}[1]{\mathbf{#1}}
\renewcommand{\c}[1]{\mathcal{#1}}
\renewcommand{\u}{\uparrow}
\renewcommand{\d}{\downarrow}
\newcommand{\be}{\begin{equation}}
\newcommand{\ee}{\end{equation}}
\newcommand{\bsigma}{\boldsymbol{\sigma}}
\newcommand{\blambda}{\boldsymbol{\lambda}}
\newcommand{\Tr}{\mathop{\mathrm{Tr}}}
\newcommand{\sgn}{\mathop{\mathrm{sgn}}}
\newcommand{\sech}{\mathop{\mathrm{sech}}}
\newcommand{\diag}{\mathop{\mathrm{diag}}}
\newcommand{\Pf}{\mathop{\mathrm{Pf}}}
\newcommand{\half}{{\textstyle\frac{1}{2}}}
\newcommand{\sh}{{\textstyle{\frac{1}{2}}}}
\newcommand{\ish}{{\textstyle{\frac{i}{2}}}}
\newcommand{\thf}{{\textstyle{\frac{3}{2}}}}
\newcommand{\SUN}{SU(\mathcal{N})}
\newcommand{\N}{\mathcal{N}}

\newcommand{\AK}[1]{[AK: #1]}
\newcommand{\KJ}[1]{[KJ: #1]}

\title{The holographic dual of an EPR pair has a wormhole}

\author{Kristan Jensen$^1$ and Andreas Karch$^2$}

\affiliation{$^1$: Department of Physics and Astronomy, University of Victoria, Victoria, BC V8W 3P6, Canada \\
$^2$: Department of Physics, University of Washington, Seattle, WA
98195-1560, USA}

\date\today

\begin{abstract}
We construct the holographic dual of two colored quasiparticles in maximally supersymmetric Yang-Mills theory entangled in a color singlet EPR pair. In the holographic dual the entanglement is encoded in a geometry of a non-traversable wormhole on the worldsheet of the flux tube connecting the pair. This gives a simple example supporting the recent claim by Maldacena and Susskind that EPR pairs and non-traversable wormholes are equivalent descriptions of the same physics.
\end{abstract}

\pacs{11.25.Tq,
03.65.Ud
}

\maketitle

{\bf Introduction:} Quantum entanglement is one of the most perplexing consequences of
quantum mechanics. Two particles, created in an entangled Einstein-Podolsky-Rosen (EPR) pair, are tied together by this famous ``spooky action at a distance". While the measurement of the properties of the pair as a whole comes with the usual quantum uncertainty, measurement of any one of the two forces the wavefunction of the system into a particular eigenstate of the measured observable completely determining the state of the EPR partner.

\begin{figure}
\includegraphics[scale=.15]{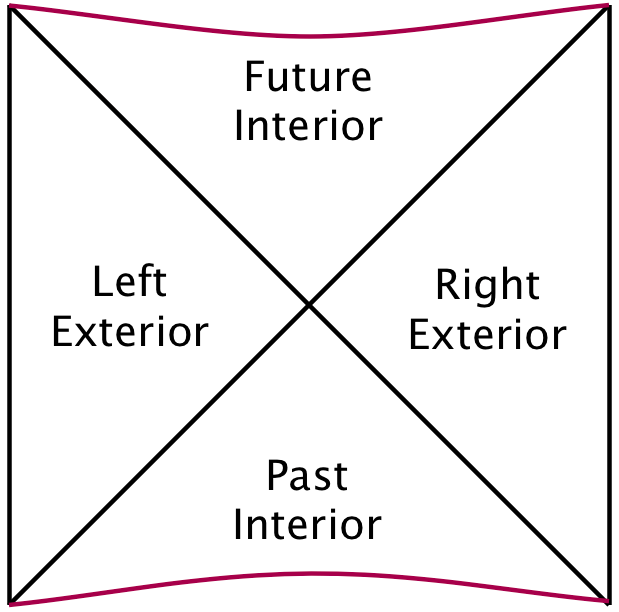}
\caption{\label{F:penrose} The Penrose diagram of an eternal AdS black hole.}
\end{figure}

Recently it has been argued by Maldacena and Susskind \cite{Maldacena:2013xja} (MS) that quantum entanglement is secretly tied to another, so far purely theoretical, construction: non-traversable wormholes, or Einstein-Rosen (ER) bridges. Their basic argument was built around the quantum mechanics of black holes. The standard Schwarzschild metric for a spherically symmetric black hole describes not just one, but two asymptotically flat black holes. The outside of the black holes are not in direct contact with each other: it is impossible to send a signal between two outside observers, Alice and Bob, through the black hole. The two outside observers can however influence what the other one observes when jumping into the black hole. Signals sent by Alice into her black hole can determine what Bob will experience after traversing the horizon and vice versa.\footnote{MS were motivated in part to seek a resolution of the AMPS firewall paradox~\cite{Almheiri:2012rt}. MS claim that this observation clarifies the crux of the AMPS paradox: what Bob perceives after falling through his horizon depends on what Alice (more generally quantum gravity in her causal wedge) sends his way.}  These are the characteristic features of an ER bridge. MS argue that the geometry of the ER bridge is the geometric manifestation of the entanglement between the two black holes. Their argument may be made more precise in the context of eternal black holes in AdS (we reproduce the Penrose diagram for such a black hole in Fig.~\ref{F:penrose}),  where we may bring the power of AdS/CFT~\cite{Maldacena:1997re} to bear.

MS go on to consider more complicated scenarios. One could have produced the entangled black hole pair by pair production, very much like a more conventional EPR pair. Alternatively, an outside observer of a one-sided black hole formed by collapse could have captured the Hawking radiation emitted by the black hole over a period of time and then
collapsed the photons into their own black hole. Clearly the original Hawking radiation
is entangled with the black hole that emitted it.

While all these examples involved entanglement between black holes as well as macroscopic wormholes, MS argued that entanglement in general should be associated with wormhole formation. Individual Hawking quanta are claimed to be connected to the black hole interior via Planck-scale wormholes encoding the entanglement. When collapsing the Hawking radiation into a second black hole, all these micro-wormholes combine into a single macroscopic ER bridge.

At first sight, such a claim sounds preposterous. Quantum entanglement is a property of any quantum mechanical system, even when gravity is absent. Why microscopic wormholes should play a role in non-gravitational systems is far from obvious. What we will demonstrate in this work is that, in fact, a single EPR pair in maximally supersymmetric Yang-Mills theory (SYM) in the limit of a large number of colors and large 't Hooft coupling has an equivalent description in terms of an ER bridge. In this limit SYM has a dual holographic description~\cite{Maldacena:1997re} in terms of string theory on asymptotically anti de-Sitter (AdS) space. In this dual description, the EPR pair formed by a single (external) quark anti-quark pair is described by a single string connecting the two quasiparticles.\footnote{A somewhat similar description of an EPR pair in terms of an AdS/BCFT construction has recently been given in~\cite{Fujita:2011fp}.} We find that there is an ER bridge on the string worldsheet. The two horizons shielding the wormhole from the two asymptotic regions carry an entropy, which can be identified with the entanglement entropy of the pair. EPR and ER, at least in this theory, can indeed be seen as two equivalent descriptions of one and the same physical reality.

{\bf The holographic EPR pair:}
A single heavy test quark in SYM is holographically dual to a fundamental string stretching from the Poincare patch horizon to the boundary of AdS. The string endpoint on the boundary represents the quark. The action of the fundamental string is proportional to $(\alpha')^{-1} \sim \sqrt{\lambda}$ where $\lambda$ is the 't Hooft coupling. Consequently its free energy, energy, and entropy are all proportional to $\sqrt{\lambda}$. Maybe most interestingly, a single quark has a zero temperature entropy of $S=\sqrt{\lambda}/2$ \cite{Karch:2008uy}. Clearly these are not the properties of a single quark in free SYM theory. In the strongly coupled SYM the quark is really a colored quasiparticle formed by the heavy test quark surrounded by a cloud of order $\sqrt{\lambda}$ gluons.

A color neutral state can be formed by making a quark anti-quark ($q$-$\bar{q}$) pair. By forcing the $q$-$\bar{q}$ state to be a color singlet, we automatically demand that the two quasiparticles are entangled. In terms of the dual string, such color singlet ``meson" like states are described by a single open string with both endpoints at the boundary. String configurations dual to a separating $q$-$\bar{q}$ had first been numerically constructed in~\cite{Herzog:2006gh,Chernicoff:2008sa}. Similar numerical solutions for light quarks, this time dual to falling strings, have been obtained in~\cite{Chesler:2008wd}.

The string connecting the quasiparticles is the holographic dual of the color fluxtube between the two. Unlike in a confining theory, in SYM the fluxtube does not give rise to a linear potential. The force between the $q$-$\bar{q}$ falls off with a Coulombic $1/r$ as demanded by the scale invariance of that theory. The $q$-$\bar{q}$ can separate arbitrarily far despite the flux between them. The flux tube connecting the two however enforces that the $q$-$\bar{q}$ is entangled into a color singlet state.

\begin{figure}
\begin{center}
\includegraphics[scale=0.31]{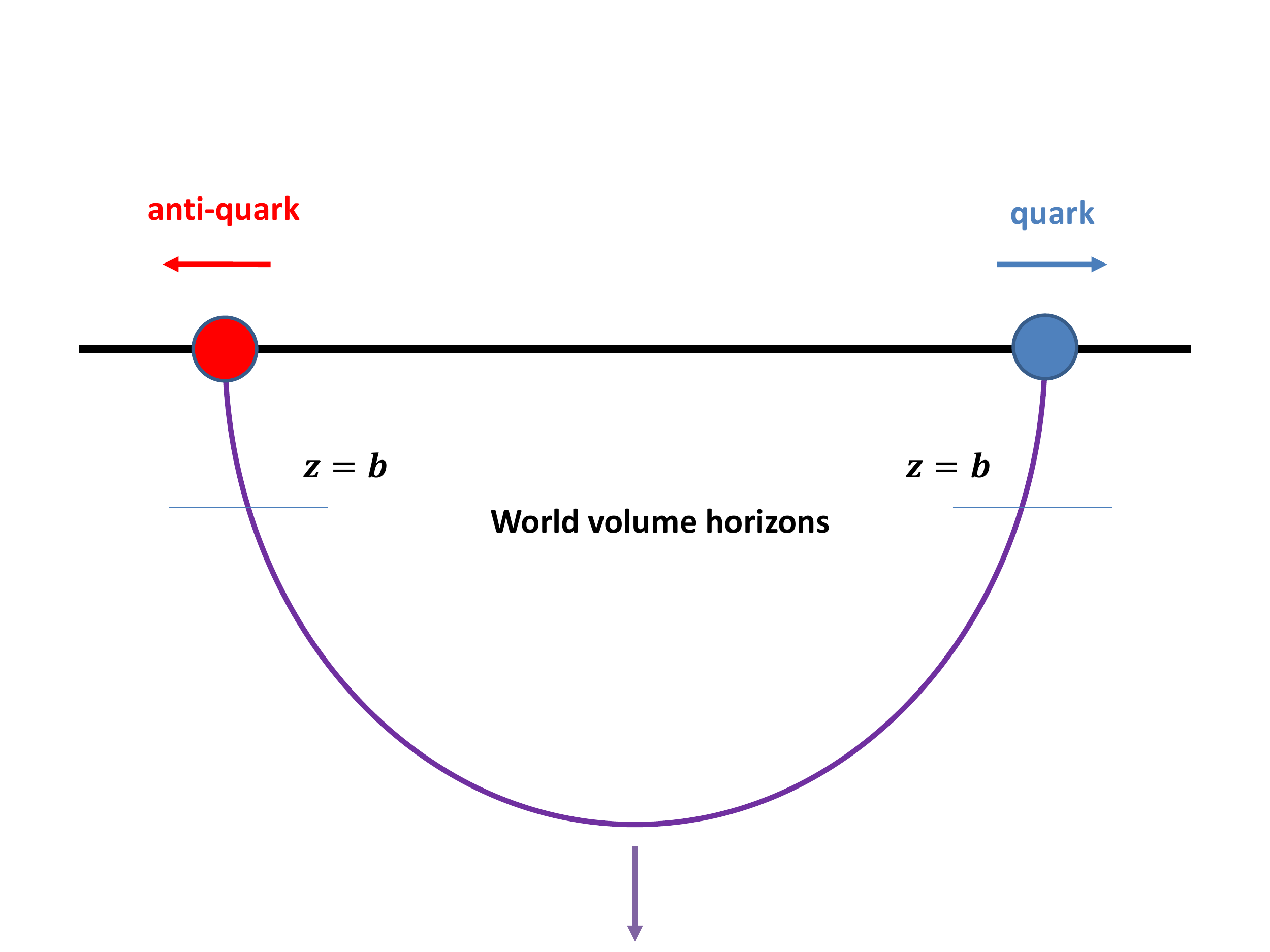}
\caption{\label{q-qbarpair}
The holographic $q$-$\bar{q}$ system entangled into a color-neutral EPR pair.}
\end{center}
\end{figure}

An analytic solution for an accelerating $q$-$\bar{q}$ pair was found in~\cite{Xiao:2008nr}. The geometry of this solution is depicted schematically in Fig.~\ref{q-qbarpair}.
In this solution the quark and anti-quark are accelerated so that their velocity asymptotically approaches the speed of light. This is a crucial feature, insofar as the two entangled particles are out of causal contact. This is a property this system shares with the original EPR pair. No signal emitted from particle Alice can reach particle Bob in a finite amount of time. It is exactly in this situation that MS claimed that entanglement is encoded in the geometry of an ER bridge.

Using Poinc\'are patch coordinates in which the background AdS$_5$ metric is
\be
ds^2 = \frac{R^2}{z^2} \left ( - dt^2 + d \vec{x}^2 + dz^2 \right )\,,
\ee
the embedding of the string is given by the expanding semicircle
\be
x^2 = t^2 +b^2 - z^2 \,.
\ee
The quark and anti-quark are located at $x=\pm \sqrt{t^2+b^2}$, accelerating away from each other for all time. They initially travel toward each other, until they turn around at $t=0$ at $x=\pm b$, then fly away from each other. At late times they go to $x=\pm\infty$ near the speed of light. For infinitely heavy test quarks, one can simply prescribe their trajectory as an external boundary condition. If the quarks are very heavy dynamical objects, the string needs to end on a flavor probe brane~\cite{Karch:2002sh} at a small but finite $z_m<b$, which is related to the quark mass by $m = \sqrt{\lambda}/(2 \pi z_m)$. In this case, the boundary conditions on the string require a constant electric field $E=m/b$ on the flavor brane. This electric field is responsible for accelerating the quasiparticles.

\begin{figure}
\begin{center}
\includegraphics[scale=0.85]{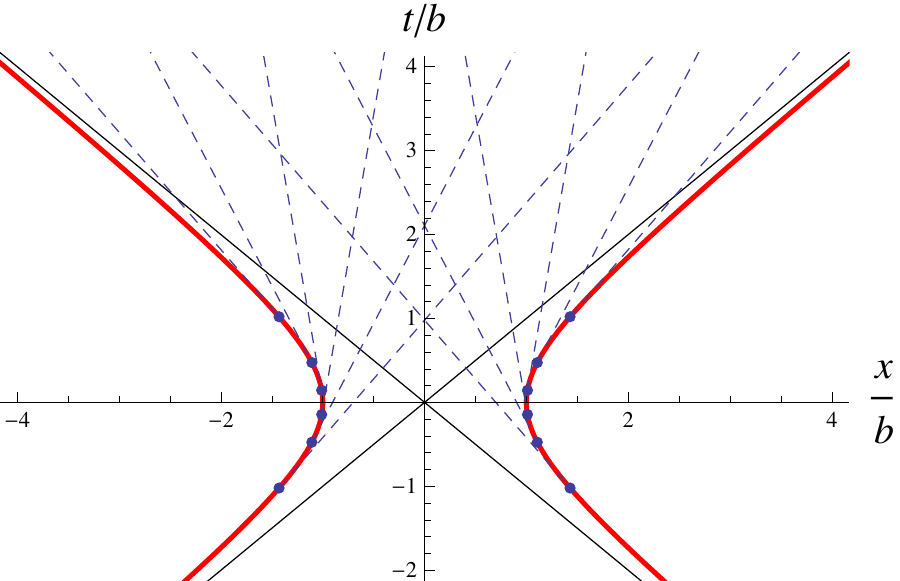}
\caption{\label{causal}
The casual structure on the string worldsheet. The thick red lines indicate the worldlines of the quark and anti-quark, and the string worldsheet fills the universe in between. The solid lines indicate the worldsheet horizons, which happens to be the location of the Rindler horizons for each of the quarks. The solid dots denote events where light rays are emitted from the quark and anti-quark into the dual worldsheet, and the dotted lines indicate the resulting lightlike trajectories. The string worldsheet clearly has the same causal structure of an eternal AdS black hole as in Figure~\protect\ref{F:penrose}.}
\end{center}
\end{figure}

The most important aspect of the worldsheet metric for us is that it has two horizons located at $z=b$, as indicated in Fig.~\ref{q-qbarpair}. To understand the causal structure on the string worldsheet, we have mapped out lightlike geodesics in the two-dimensional universe living on the string worldsheet in Fig.~\ref{causal}. One look at the picture shows that this causal structure is identical to that of the eternal AdS black hole pictured in Fig.~\ref{F:penrose}. The holographic dual of the EPR pair then has two horizons and an Einstein-Rosen bridge connecting them. By an ER bridge here we simply refer to a geometry with spacelike paths connecting causally disconnected regions on the worldsheet.

For completeness, we elaborate briefly on the causal structure. All lightlike geodesics on the worldsheet hit either the left anti-quark or right quark exactly once at a time we denote as $t_0$. In terms of $t_0$, the worldsheet light rays are
\be
x = \pm\frac{t_0t+b^2}{\sqrt{t_0^2+b^2}}\,,
\ee
where we take the plus (minus) sign for a light ray which hits the right (left) quark. One might worry that the causal structure one obtains from the worldsheet is not quite the causal structure one obtains from the bulk, pulled back to the worldsheet. That is, perhaps we could send information to a point on the worldsheet via bulk fields faster than we could by sending the signal along the string. However, a quick calculation shows that these worldline geodesics are precisely the bulk lightlike geodesics that propagate in the $(t,x,z)$ directions.

At least in this particular example we have demonstrated that the physics of entanglement in a single EPR pair is equivalently captured by the geometry of an ER bridge. The fact that the worldsheet causal structure is inherited from the bulk causal structure makes it easy to see what is happening in
a more general scenario. As long as the $q$-$\bar{q}$ pair separates fast enough to ensure that the pair loses causal contact, we will inherit two worldsheet horizons connected by an ER bridge from the bulk Rindler horizons of the accelerated charges. For other solutions, e.g. where the pair oscillates around the center of mass, the causal structure on the worldsheet is trivial\footnote{For the closely related case of a holographic $q$-$\bar{q}$ pair moving in global AdS the worldsheet casual structure of the general solution has been worked out in~\cite{Chernicoff:2010yv}.}.

{\bf Global perspective:}
There is a thermal entropy associated with the worldsheet horizons. We would like to understand the relation between this thermal entropy and the entanglement of the $q$-$\bar{q}$ EPR pair. To do so, we make an intermediate step and show that the accelerating string of~\cite{Xiao:2008nr} can be written as a static string stretching straight across global AdS.  The horizon entropy and entanglement may be easily understood in this global picture. Recall that global AdS is the geodesic completion of the Poinc\'are patch. Its boundary is time cross a spatial sphere. To go from global AdS to the Poinc\'are patch, we represent AdS$_{d+1}$ as a hypersurface in one higher dimension satisfying
\be
-(X^0)^2 - (X^{d+1})^2 + \sum_{i=1}^d X^i X^i = -1\,.
\ee
We can get global AdS by defining
\be
\nonumber
X^0 = \cosh \rho \sin \tau\,, \hspace{.2cm}
X^{d+2} = \cosh \rho \cos \tau\,, \hspace{.2cm}
X^i = \sinh \rho \,e^i\,,
 \ee
where $e^i$ are vectors on the $\mathbb{S}^{d-1}$ unit sphere. These coordinates cover the entire hypersurface. Now consider the static string stretched in the $\rho$ direction (and evolving in time $\tau$), from one pole of the boundary sphere to the other:
\be
X^1 = \pm \sinh\rho \cos\theta_0\,, \hspace{.2cm}
X^2 = \pm \sinh\rho \sin\theta_0\,, \hspace{.2cm}
X^{i>2} = 0\,.
\ee
Next we restrict to the Poinc\'are patch, defining
\begin{align*}
X^{1,d+1} &= \frac{\pm 1-z^2 + t^2 - x^2 -\vec{y}^2}{2z}\,,
\\
X^0 &= \frac{t}{z}\,, \hspace{.2cm}
X^2 = \frac{x}{z}\,, \hspace{.2cm}
X^{3, \ldots, d} = \frac{\vec{y}}{z}\,.
\end{align*}
The string sits at $\vec{y}=\vec{0}$, so that points on the string satisfy
\be
\nonumber
\frac{X^1}{X^2} = \cot(\theta_0) = \frac{1-z^2 + t^2 - x^2}{2x}.
\ee
Solving for $x$ gives (up to a constant shift in $x$) the expanding string of~\cite{Xiao:2008nr}.

We would now like to quantify the entanglement of our $q$-$\bar{q}$ pair. This is a somewhat perilous task. In a conventional EPR pair of two entangled spins, one may trace over only the spin degrees of freedom for one particle and thereby construct a reduced density matrix encoding the spin entanglement. Our setup is rather different. The entanglement is dominated by the large $N$ color correlations between our quasiparticles so that we have a color singlet. We cannot trace over just the color states of a colored quasiparticle (which includes a cloud of $\mathcal{N}=4$ fields surrounding the test particle), and moreover there is no gauge-invariant notion of a reduced density matrix encoding color entanglement. Thus we must employ other observables to describe the entanglement. In what follows we will content ourselves with position-space entanglement entropy (EE) for regions surrounding the quasiparticles. Unfortunately, these EEs do not a priori cleanly separate into a piece describing the pair plus other contributions. Despite all of these difficulties, our results suggest that the position-space EE is indeed describing the entanglement of the $q$-$\bar{q}$ pair.

For the straight string in global AdS it is straightforward to calculate the EE for certain regions around a single quasiparticle using a generalization~\cite{andykristan} of the trick developed in~\cite{Casini:2011kv}. Writing global AdS in yet another coordinate system, hyperbolic slicing, the Ryu-Takayanagi entangling surface~\cite{Ryu:2006bv} maps to a genuine horizon with a thermal entropy. The method of~\cite{andykristan,Casini:2011kv} shows that this entropy is equal to the EE one gets from tracing out everything below a line of constant latitude $\theta$ on the boundary sphere (the $q$-$\bar{q}$ reside at the north and south poles). By computing the string's contribution to the thermal entropy we obtain the EE after tracing out the anti-quark and everything else below $\theta$ to be
\be
\label{E:quarkEE}
S_{EE} = \frac{\sqrt{\lambda}}{3}\,,
\ee
notably independent of the latitude. In fact, using the same hyperbolic slicing of the Poinc\'are patch, the author of~\cite{Xiao:2008nr} showed that the horizon has a temperature $T=1/(2\pi b)$. From here it is easy to compute that the thermal entropy of the worldsheet horizon is also $\sqrt{\lambda}/3$. Furthermore, the trick of~\cite{Casini:2011kv} shows that not only are the thermal and entanglement entropies equal, but the entire thermal and reduced density matrices are too.

We will argue that~\eqref{E:quarkEE} describes the entanglement of the EPR pair. To do so, we find it useful to compute the position-space EEs in a variety of configurations for the $q$-$\bar{q}$ pair. First, consider a configuration in which the $quark$ uniformly accelerates along the worldline $x(t)=\sqrt{b^2+t^2}$ as before, but the $\bar{q}$ travels along an arbitrary path inside the left Rindler wedge given by $x\pm t < 0$. Now trace over the left Rindler wedge. The $q$ and $\bar{q}$ are out of causal contact for all time, and so the reduced density matrix on the right Rindler wedge does not depend on the path taken by the anti-quark. As a result, the string's contribution to the EE of the right Rindler wedge is still~\eqref{E:quarkEE}. Since the theory is in the vacuum, the EE of the left wedge is equal to that of the right wedge and so is also~\eqref{E:quarkEE}.

Our second example involves an entangled $q$-$\bar{q}$ pair which uniformly accelerate in the same Rindler wedge, say the right one. Suppose that their accelerations are nearly identical, that is they travel along nearby paths $x_i(t) = \sqrt{b_i^2+t^2}$ with $b_1\sim b_2$. Now trace over the left Rindler wedge. The method of~\cite{andykristan,Casini:2011kv} shows that the EE is equal to the thermal entropy on hyperbolic space, where the $q$-$\bar{q}$ are static and near each other. In the holographically dual geometry, the dual string hangs in a U-shape from one point on the boundary to another over a hyperbolic horizon. Clearly the string's contribution to the EE will be rather different in this scenario. Since the theory is in the vacuum, the EE of the right Rindler wedge is equal to that of the left wedge, which may be computed more easily by~\cite{andykristan,Casini:2011kv}: it is still the thermal entropy of the dual hyperbolic black brane, but in this case the dual string connecting the $q$-$\bar{q}$ pair is behind the horizon. As a result the string's contribution to the EE is \emph{zero}.

Putting these results together, we see that the correction to the EE~\eqref{E:quarkEE} is robust. It holds for a wide family of configurations in which we trace over a region containing a single colored quasiparticle. Moreover, in our example with the static string in global AdS the correction~\eqref{E:quarkEE} was independent of the latitude of the entangling surface. As a result,~\eqref{E:quarkEE} appears to characterize the entanglement of the $q$-$\bar{q}$ pair as claimed, rather than the entanglement of the $q$ and $\bar{q}$ with the $\mathcal{N}=4$ fields.

Both the expanding string and the static string in global AdS lead to the same EE, but the expanding string has horizons and an ER bridge while the static string has neither. The primary difference between these solutions is that the expanding $q$-$\bar{q}$ are out of causal contact, while the static $q$-$\bar{q}$ pair are in causal contact. This is completely consistent with the claim of MS, who do not say that entanglement in general is geometrized into a wormhole, but only entanglement between causally disconnected degrees of freedom.

To summarize, the worldsheet horizons of the expanding string carry a thermal entropy of $\sqrt{\lambda}/3$. We have argued that this is also the EE one would get from tracing over the degrees of freedom of a single quark placed on the south pole of a sphere (or~\cite{andykristan,Casini:2011kv} equivalently, the EE of a spherical region around a single quark). In this sense our $q$-$\bar{q}$ pair is maximally entangled like an ordinary EPR pair.

Note that while the entanglement entropy nicely accounts for the physics of the EPR pair, it does not completely account for the thermal entropy of a static quasiparticle in flat space. In general AdS$_{d+1}$ the spherical EE of this quark is $d/[2(d-1)]$ times the zero temperature limit of the thermal entropy. Nevertheless, the latter can still be extracted from the entanglement of the EPR pair. As explained in~\cite{Galante:2013wta} we can get the thermal entropy of a single quasiparticle in flat space from an $n \rightarrow 0$ limit of the Renyi entropies $S_n$, which one may calculate from hyperbolic foliation of AdS~\cite{Hung:2011nu}.

{\bf Correlations:} What are the physical consequences of the ER bridge? To answer this question it is helpful to return to the EPR paradox and entanglement in field theory. In a conventional EPR pair, quantum entanglement is manifest through the violation of Bell's inequalities. The latter may be understood as inequalities obeyed by the equal-time spin-spin correlator upon demanding local realism. However, a much simpler diagnostic of entanglement is a non-vanishing \emph{connected} two-point function at spacelike separation. Local realism would demand that the connected piece of e.g. the equal-time spin-spin correlator vanishes. In the conventional EPR pair, the connected spin correlations may be used to quantify the spin entanglement, However, in a general context like our entangled $q$-$\bar{q}$ pair, these connected correlations only signal the \emph{existence} of entanglement.

Returning to our holographic system, we note that any connected string worldsheet, with or without worldsheet horizons, will give to rise non-vanishing $O(\sqrt{\lambda})$ connected correlations signaling entanglement. What is special about the worldsheet with the wormhole geometry is that, due to the horizons, all causal Green's functions between the quark and anti-quark vanish. This is expected in the field theory, as the quark and anti-quark are always spacelike separated. The ER bridge on the worldsheet however still allows connected correlations between the $q$-$\bar{q}$ pair, thereby signaling entanglement. Conversely, the entanglement of the causally disconnected EPR pair should result in connected $O(\sqrt{\lambda})$ correlations between the quark and anti-quark. These spacelike correlations are only possible when the dual string has a wormhole connecting the causally disconnected regions near the string endpoints.

Note that for a disconnected configuration of two strings, each of which has one endpoint at the boundary and the other end reaching behind the horizon, the connected two-point correlation function of any operator dual to a string fluctuation vanishes at order $\sqrt{\lambda}$. Such correlators only receive contributions from bulk graviton exchange, which is suppressed by a factor of Newton's constant, which is small (it is parametrically $O(1/N^2)$ in the present instance) for holography to be applicable.

{\bf Discussion:} We have been able to demonstrate that the holographic dual of a single EPR pair in strongly coupled SYM has the geometry of an ER bridge. This takes a step towards establishing the conjecture of MS that even for single pairs entanglement can be equivalently viewed as formation of a microscopic wormhole geometry. In our construction this wormhole lives on the string connecting the quark and anti-quark in the dual AdS spacetime. While our construction does not rely on a macroscopic wormhole geometry, our analysis requires a classical worldsheet description. This limit is reliable at large $\lambda$, when our quark quasiparticle has a large number $O(\sqrt{\lambda})$ of internal degrees of freedom due to the gluon cloud surrounding the quark. Our holographic EPR pair should give an ideal laboratory for further theoretical studies of the connection between entanglement, ER bridges, and perhaps even firewalls~\cite{Almheiri:2012rt,Braunstein:2009my}. For instance, imagine starting with a $q$-$\bar{q}$ pair at rest, accelerating one quark away from the other, then bringing it to rest a long time later. The holographically dual worldsheet description will involve two-sided horizon formation, followed by evaporation.

In concluding this Letter we comment on the relation between entanglement and wormholes. Our accelerating EPR pair comes from a static $q$-$\bar{q}$ pair on $\mathbb{S}^3$ with the quarks located at antipodal points. Both are EPR pairs: tracing out one quark leads to the same entanglement in each case. However, the dual worldsheet of the accelerating pair has an ER bridge, while the dual of the static pair does not. Earlier, we pointed out that this is consistent with the claim of MS (who say that an ER bridge should encode entanglement between causally disconnected degrees of freedom). Moreover the wormhole is required to mediate the connected spacelike correlations between the $q$ and $\bar{q}$ which signal entanglement. In this sense, the critical ingredient in our system is the existence of spacelike regions connecting the causally disconnected regions where our entangled degrees of freedom live.

\begin{figure}
\includegraphics[scale=.20]{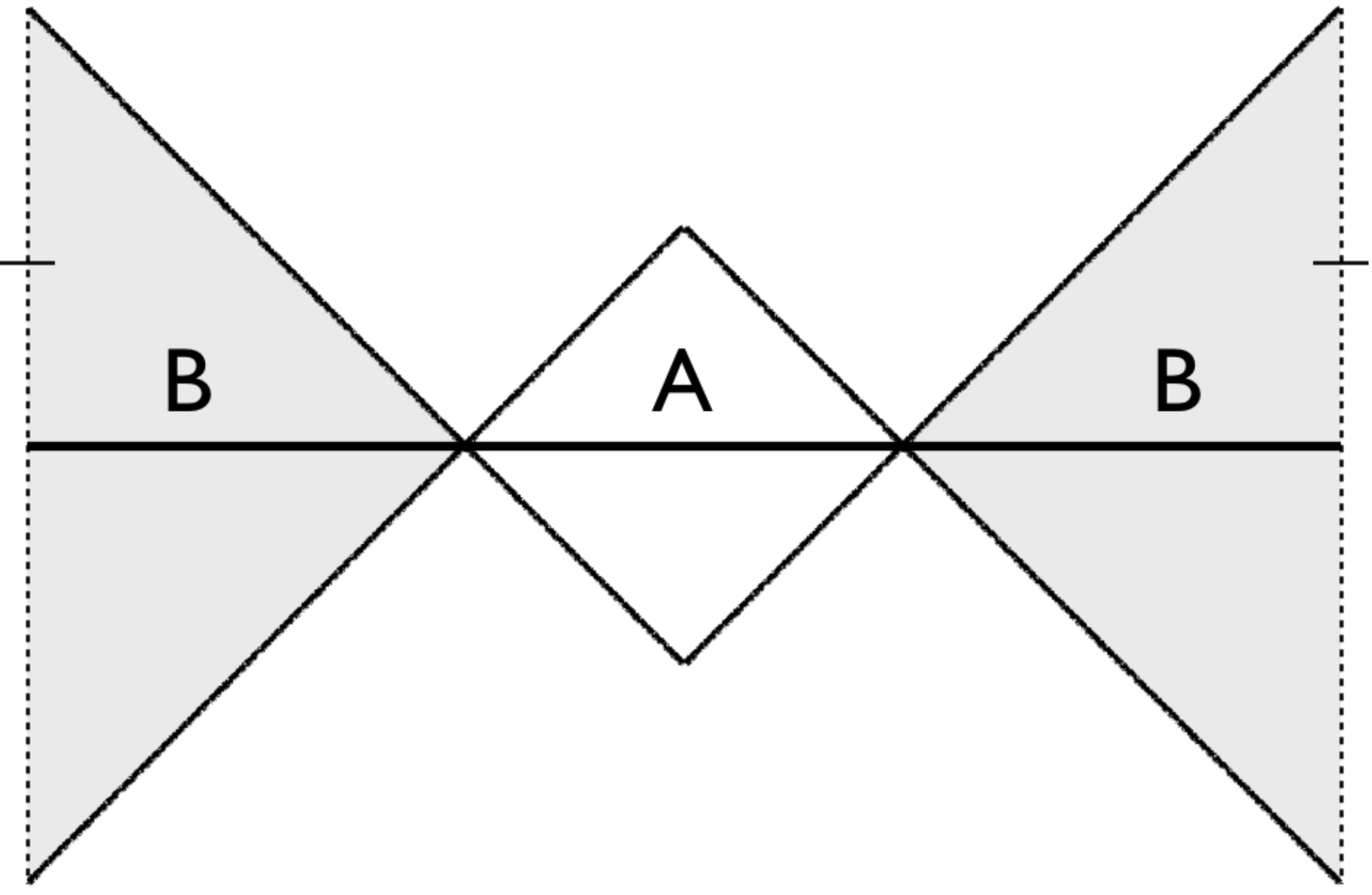}
\caption{\label{F:EE_EPR} Entanglement entropy for field theory on a circle, with ends identified. By tracing over the degrees of freedom in the causal development of B (the shaded region) we obtain a reduced density matrix on A which completely specifies the state of the system within its causal development.}
\end{figure}

Note that when defining position-space EE in field theory, one is also dividing up spacetime into causal wedges which necessarily have causal horizons. Position-space EE measures EPR-like correlations between different causal ``universes'' by construction. We depict this division for a field theory on a circle in Fig.~\ref{F:EE_EPR}, where we divide the circle into regions $A$ and $B$ and trace over $B$. The resulting entanglement is a property of the causal developments of $A$ \emph{and} $B$, and measures correlations between the two. In that sense entanglement involves causal horizons (and so spacelike trajectories connecting them) almost as a matter of definition. For the simplest case of the ER bridge between two Rindler horizons, the connection between causal wedges and entanglement was recently emphasized in~\cite{Czech:2012bh,Czech:2012be}. Our holographic construction makes this connection quantitative. The worldsheet horizon does not simply encode the fact that there is entanglement and tells us what degrees of freedom need to be traced over, it also gives us a quantitative way to calculate the amount of entanglement entropy from the horizon properties.

\section*{Acknowledgements} We would like to thank S.~Minwalla, A.~Nelson, and A.~Ritz, and especially A.~Yarom for useful discussions. We also thank J.~Maldacena, T.~Takayanagi, and  M.~van Raamsdonk for comments on an earlier version of this manuscript. AK would also like to thank the KMI at Nagoya University for hospitality during the final stages of this work. The work of KJ was supported by NSERC, Canada, while the work AK has been supported in part by the U.S. Department of Energy under Grant No.~DE-FG02-96ER40956.

\bibliography{EPR}

\end{document}